# *De novo* topology optimization of Total Ossicular Replacement Prostheses

Mario Milazzo[a,b], Pieter G. G. Muyshondt[c], Josephine Carstensen[a], Joris J. J. Dirckx[c], Serena Danti[a,b,d], Markus J. Buehler[a,*]

[a] Dept. of Civil and Environmental Engineering at Massachusetts Institute of Technology, 77 Massachusetts Ave, Cambridge, MA 02139, USA
[b] The BioRobotics Institute, Scuola Superiore Sant'Anna, Viale Rinaldo Piaggio 34, 56025 Pontedera (PI), Italy
[c] Laboratory of Biophysics and Biomedical Physics, University of Antwerp, Groenenborgerlaan 171, B-2020 Antwerp, Belgium
[d] Dept. of Civil and Industrial Engineering, University of Pisa, Largo L. Lazzarino 2, 56122 Pisa, Italy

*\* Correspondence should be addressed to:*

Prof. Markus J. Buehler
Laboratory for Atomistic and Molecular Mechanics (LAMM)
Department of Civil and Environmental Engineering
Massachusetts Institute of Technology, 77 Massachusetts Ave., Cambridge, Massachusetts 02139 (USA)
mbuehler@mit.edu, +1.617.452.2750

**ABSTRACT**

Conductive hearing loss, due to middle ear pathologies or traumas, affects more than 5% of the population worldwide. Passive prostheses to replace the ossicular chain mainly rely on piston-like titanium and/or hydroxyapatite devices, which in the long term suffer from extrusion. Although the basic shape of such devices always consists of a base for contact with the eardrum and a stem to have mechanical connection with the residual bony structures, a plethora of topologies have been proposed, mainly to help surgical positioning. In this work, we optimize the topology of a total ossicular replacement prosthesis, by maximizing the global stiffness and under the smallest possible volume constraint that ensures material continuity. This investigation optimizes the prosthesis topology in response to static displacement loads with amplitudes that normally occur during sound stimulation in a frequency range between 100 Hz and 10 kHz. Following earlier studies, we discuss how the presence and arrangement of holes on the surface of the prosthesis plate in contact with the umbo affect the overall geometry. Finally, we validate the designs through a finite-element model, in which we assess the prosthesis performance upon dynamic sound pressure loads by considering four different constitutive materials: titanium, cortical bone, silk, and collagen/hydroxyapatite. The results show that the selected prostheses present, almost independently of their constitutive material, a vibroacoustic behavior close to that of the native ossicular chain, with a slight almost constant positive shift that reaches a maximum of ≈5 dB close to 1 kHz. This work represents a reference for the development of a new generation of middle ear prostheses with non-conventional topologies for fabrication via additive manufacturing technologies or ultraprecision machining in order to create patient-specific devices to recover from conductive hearing loss.

# 1. INTRODUCTION

Conductive hearing loss, due to malfunctioning or traumas of the middle ear, is a relevant clinical issue that affects more than 5% of the population worldwide. The function of the middle ear is to transmit sounds as collected by the auricle (i.e., outer ear) up to the cochlea (i.e., inner ear), where amplification and mechanical-to-electrical conversion of the sound signals takes place (Neely, 1998). The tympanic membrane separates the outer form the middle ear compartments and prime the mechanical function of the ossicles. Defective ossicular bones are replaced with material-based prostheses under a surgical technique known as ossiculoplasty. Middle ear passive implants thus restore the mechanical connection between the eardrum and the ossicular chain (Merchant and Rosowski, 2013) (Fig. 1).

Based on the specific clinical need, two main approaches are reported to restore conductive hearing loss by passive middle ear implants: (1) Autografts via incus repositioning (gold standard), if possible, which involves incus reshaping during surgery – Fig. 1B (Milazzo et al., 2018); (2) Preformed ossicular replacement prostheses (ORPs). Partial ORPs (PORPs) are employed when the stapes is still intact (Fig. 1C), and total ORPs (TORPs) otherwise (Fig. 1D) (Milazzo et al., 2016; Saliba et al., 2018; Ziąbka and Malec, 2019). These devices can be made of different materials, including bone and cartilage allografts, or synthetic materials. In the latter case, and with titanium in particular, tissue grafts are needed to support the tympanic membrane interface, namely cartilaginous slice from tragus or perichondrium. If the eardrum is damaged, tissue grafts, in this case tragal cartilage and fascia, are also applied to patch or replace it (Fig. 1E).

The first attempt to reconstruct the ossicles under the approach (3) was carried out by Shea in 1976 (Shea, 1976). Following this, clinicians have employed autogenous and allogenic materials, as they offer of-the-shell preformed devices of more practical use than sculpted grafts. Although approach (2) is still considered the gold standard, it is not always possible. Allograft materials have been replaced by synthetic (known as alloplastic) materials, such as titanium, alloplastic polymers (e.g., Teflon, porous polyethylene), and even hydroxyapatite that well mimics the mineral component of bone (Kapur and Jayaramchandran, 1992; Smyth, 1982). Titanium has been the most widely used material for its efficiency and specific material properties (i.e., low specific density, durability, resistance to corrosion, as well as MRI compatibility (Alaani and Raut, 2010; Kwok et al., 2003; Mardassi et al., 2011; Martin et al., 2005)). Its first successful use was documented in the nineties (Stupp et al., 1996), and from the beginning of this century, clinicians have massively employed it to the detriment to all other approaches, by interposing a slice of cartilage between the prosthesis and the eardrum to limit the extrusion phenomena (Gardner et al., 2004; Javia and Ruckenstein, 2006). Lately, researchers have presented new interesting designs with adjustable lengths or specific joints than can be tuned *in situ*, which preserve the global piston-like shape. Concerning the acoustic performance, none statistically significant improvement have been presented. Although these new prosthetic designs help the positioning inside the middle ear cavity, they still need the interposition of the cartilaginous buffer (Arechvo et al., 2012; Beleites et al., 2007; Gostian et al., 2013; Gottlieb et al., 2016; Kuru et al., 2015; Yamada and Goode, 2010; Zhao et al., 2005).

The materials to be used for middle ear replacements should take into account durability and mechanical performance within an often inflamed and contaminated environment, which concurs to cause implant failure(Charlett et al., 2007; Mangham, 2008; Shah et al., 2013). The middle ear compatibility of the non-bioresorbable synthetic prostheses is therefore limited, with extrusion rates reported to be up ≈ 20% after 5 years even when interposing a slice of cartilage (Nguyen et al., 2004), eventually enhancing unwanted device migration. Extrusion rate, however, depends on material type, surgical techniques and expertise of the otologic surgeon. To improve prosthesis durability by controlling the phenomena at the prosthesis/biological interfaces, researchers have pursued two main approaches: developing prostheses made of (1) bulk synthetic materials with antimicrobial additives (e.g., silver nanoparticles (Danti et al., 2010; Ziabka et al., 2019; Ziąbka and Malec, 2019)) or (2) engineered biological materials (Berrettini et al., 2011; Danti et al., 2009; Milazzo et al., 2016). These studies have mainly focused on solving the middle ear compatibility problem, or developing new methodologies to fabricate miniature prostheses with specific features. For instance, Milazzo *et al.*

did a preliminary study on how holes and carved motifs on the prosthesis umbo plate not only reduce the weight, but also allow a better positioning of the prosthesis during surgery, limiting the migration effect (Milazzo et al., 2016).

To the best of the authors' knowledge, no evidence-based studies exist that aim to optimize the geometry of ossicular replacements without taking into account the limitations of manufacturability. Due to the limits of traditional fabrication techniques, researchers have focused on changing material properties or adding features, without attempting to improve the traditional piston-like geometry of the prostheses.

The advent of additive manufacturing and ultraprecision machining enables the fabrication of complex geometries that are generally not achievable using traditional technologies, employing both biological and synthetic materials (Milazzo et al., 2019). With these premises, we provide an objective study on a TORP geometry, aiming at maximizing its stiffness under the smallest possible volume constraint that preserves material continuity. We believe that this approach, originating from the structural design area for automotive and aircraft applications (Bendsøe and Kikuchi, 1988; Bendsøe and Sigmund, 2002), can be successfully applied to design a replacement prosthesis for the tiniest bones of the human body, namely the ossicles. By applying static displacement loads that correspond to the amplitudes of middle ear motion in the frequency range of 100 Hz–10 kHz, we investigate how the prosthesis geometry is affected by the presence of holes in the umbo plate. Finally, we assess the mechanical response of four selected optimized prosthesis geometries in a more complex simulation environment mimicking the middle ear under dynamic sound pressure loads, investigating four different materials: titanium, cortical bone, silk, and a Collagen/hydroxyapatite (HA) composite. The last two, so far not used for otological applications, represent a potential alternative to the traditional materials: silk is a bioinert material that can be properly shaped via 3D printing and/or ultraprecision machining (Li et al., 2016; Ling et al., 2016; Milazzo et al., 2019); Collagen/HA is a composite that mimics the bone tissue through its main components (Jung and Buehler, 2017; Milazzo et al., 2019; Nair et al., 2013).

## 2. MATERIALS AND METHODS

### 2.1. TOPOLOGY OPTIMIZATION

The optimization problem is formulated on an initial volume modeled with Solidworks 2019 (Dassault Systèmes, Johnston, RI, USA). Fig. 2 depicts the main dimensions of the initial global volume using different total lengths ($L$), from 5 mm to 7 mm, in order to estimate the effect of variations in prosthesis' length due to patient anatomical variability on the shape of the prosthesis. Fig. 2C1-C4 show the four initial cases with different arrangements of holes in the umbo plate: Case 1 – bulk (Fig. 2C1); Case 2 – big holes with borders of 0.2-mm thickness (Fig. 2C2); Case 3 – small holes, non-symmetrically positioned on the umbo plate (Fig. 2C3); Case 4 – small holes, symmetrically positioned on the umbo plate (Fig. 2C4). The 0.2-mm thickness is one of the requirements from doctors in order to ensure a reliable manipulability of the prosthesis during surgery. Other constraints are the dimensions of $\Omega_{OW}$ and $\Omega_u$, representing the contact surfaces with the oval window and the tympanic membrane, respectively. As for $\Omega_{OW}$, the dimensions shown in Fig. 2A (i.e., 0.7 mm × 0.3 mm) resemble the dimensions traditionally used for the connection to the stapes plate. Concerning $\Omega_u$, the area (i.e., 4 mm$^2$) is a trade-off between the actual area of the umbo (i.e., about 1 mm$^2$) and a stability requirement.

To process the optimization task, we use SIMULIA-Tosca (Dassault Systèmes, Johnston, RI, USA) to calculate the prosthesis' material distribution that maximizes the overall stiffness to enhance wave propagation along the prosthesis, while having the smallest possible prosthesis volume that ensures material continuity. As mechanical constraints, we fix the transversal displacements at $\Omega_{OW}$ and $\Omega_u$, assuming a purely piston-like motion of the prosthesis. The optimization process takes into account the application of static displacement loads on the same surfaces, as depicted in Fig. 2D-E, in order to achieve an optimized structure for all frequency cases that is compatible with all loading conditions (Hato et al., 2003). We also apply geometrical constraints to the body to be optimized (grey area in Fig. 2D) by avoiding any modification to the surfaces $\Omega_{OW}$ and $\Omega_u$ and by imposing a planar symmetry with the *x-y* plane. In this way, a simplified symmetrical structure can be achieved in view of an easier fabrication. The resulting geometries are then post-processed in MeshLab (Version 2016.12) in order to cancel out outliers among the points on the surface, to map the point cloud, to smooth the surface through a Gaussian filter and to round the main edges (Cignoni et al., 2008). The material is pure titanium (Young's modulus = 116 GPa, Poisson's ratio = 0.32, mass density = 4506 kg/m³).

## 2.2. ASSESSMENT IN A FINITE-ELEMENT MODEL

To validate the conductive performance of the optimized prosthesis designs, prosthesis geometries are implemented in an existing 3D finite-element model of the human middle ear (De Greef et al., 2017). The FE model is created in COMSOL 5.4 (COMSOL Multiphysics, Burlington, MA, USA) and simulates the linear harmonic vibration of the middle ear in response to sound-pressure stimulation of the tympanic membrane. The pars tensa of the tympanic membrane is modeled by using an orthotropic elastic description with different radial and circumferential elastic moduli. The radial Young's modulus decreases away from the umbo, as explained in (De Greef et al., 2017). All other components are modeled as isotropic materials. To simulate the cochlear load on the oval window, we make use of the cochlear impedance data measured in (Puria et al., 1997). The model with an intact ossicular chain is validated to the average of the stapes velocity measurements, normalized to the input pressure at the tympanic membrane, as established in (Rosowski et al., 2007). To study the performance of the prostheses, all structures are removed from the model except for the tympanic membrane and oval window. At the oval-window side, the stapes is cut at the crura just distal to the lateral footplate surface, remaining with only the footplate and annular ligament. The prosthesis is inserted between the umbo and footplate by manual positioning, with the prosthesis base plate facing the anterior direction (Fig. 4A-B). To match the prosthesis to the shapes of the tympanic membrane and footplate, we slightly adjust the geometries of the components: at the umbo, the tympanic membrane is slightly cut to adapt to the shape of the prosthesis plate; at the oval window, the prosthesis is cut at its distal end to adapt to the shape of the footplate. The prosthesis, tympanic membrane, and footplate share translational degrees of freedom at their connections. The prosthesis is modeled by using material properties of pure titanium (Young's modulus = 116 GPa, Poisson's ratio = 0.32, mass density = 4506 kg/m³), cortical bone (Young's modulus = 14.1 GPa, Poisson's ratio = 0.3, mass density = 2200 kg/m³), silk (Young's modulus = 4.2 GPa, Poisson's ratio = 0.3, mass density = 1390 kg/m³, (Li et al., 2016)), and a composite made of Collagen/HA (Young's modulus = 1.5 GPa, Poisson's ratio = 0.3, mass density = 1379 kg/m³, average values from (Clarke et al., 1993; Lawson and Czernuszka, 1998; Yamauchi et al., 2004)). From the model, the footplate vibration velocity level is calculated as a function of frequency in response to uniform harmonic sound pressure at the TM of 1-Pa amplitude.

## 3. RESULTS AND DISCUSSION

We investigate how the topology of middle ear prostheses can be optimized with minimal manufacturability constraints, in view of a fabrication via additive manufacturing or ultraprecision machining.

Since we are in the linear regime under the present loading conditions, it is expected that changing the material would have no significant effect on the final outcomes of the geometry. The presented results for the topology optimization are, thus, based on titanium as constitutive material.

Fig. *3*A shows the geometry of a prosthesis before and after filtering and smoothing the raw outcomes from SIMULIA-Tosca via MeshLab. The shapes obtained in this study resemble the well-known piston-like geometry of commercial devices and earlier studies, with the shaft asymmetrically positioned with respect to the umbo plate to compensate the anatomic distance between the axes of the two contact interfaces (i.e., umbo and oval window) (Milazzo et al., 2016; Ziąbka and Malec, 2019). The 5-mm devices present a simple piston-like geometry that does not significantly change with the presence of the holes. However, some novel aspects appear for the 6-mm and 7-mm prostheses, which often show two/three branches that connect the umbo plate with the ellipsoidal shaft. Between the 6-mm and 7-mm prosthesis, we observe similar results for Cases 1 to 3 with a relevant deviation for Case 4, in which we notice two flexible branches for the 7-mm device that are absent in the 6 mm and 5 mm prostheses. Ordinary loads applied to the devices during its function are expected to very low due to the small displacements (i.e., nanometer range) occurring at the interfaces. Therefore, putative residual stresses after fabrication would not cause catastrophic failure. In contrast, the highest loads on the prostheses will occur during the handling for the *in situ* placement due to the firm grip exerted by surgical tools. Nevertheless, based on the traditional surgical approach, the prostheses will be handled by the umbo footplate which was designed as the thickest part of the device, thus ensuring a proper and safe management.

To evaluate the performance of the prostheses in the middle-ear model, we use a 5-mm prosthesis made of titanium, cortical bone, silk, or Collagen/HA. This length is chosen as it is the best adapted to the dimensions of the available finite element model. In Fig. 5, we compare the footplate velocity level (referenced to 50 nm/s/Pa) of the native middle ear (Fig. 4A, Fig. 5 – green curve) and the four optimized prostheses, positioned in the model as exemplified in Fig. 4B.

In Fig. 5, we observe a general behavior for all prosthesis responses, which retrace the response of the native middle ear well. To describe the response behavior, we distinguish five common frequency (*f*) regions: *I*) $f \leq 700$ Hz: all colored curves overlap almost perfectly and are shifted by ≈+2 dB with respect to the native model; *II*) $700$ Hz $< f \leq 1500$ Hz: for all prosthesis curves, a positive deviation of ~+5 dB is noticeable close to 1 kHz and the response is shifted to higher frequencies as the constitutive material becomes less stiff; *III*) $1500$ Hz $< f \leq 2500$ Hz: we observe a strong match of the colored curves with the green curve, with absolute deviations not higher than 2 dB. The only exception is Fig. 5D – Collagen/HA, where we notice a small positive shift for Case 1; *IV*) $2500$ Hz $< f \leq 7000$ Hz: we notice a behavior similar to *I*) for Fig. 5A to Fig. 5C. Fig. 5D presents a good match of Cases 2 to 4 with the native middle ear curve, with a negative shift of ~-2 dB for Case 1; *V*) $7000$ Hz $< f \leq 10$ kHz: titanium and cortical bone present similar trends. The first two panels show a positive shift similar to *I*) and *IV*), with a slight difference for Case 1 – cortical bone, which secedes few decibels from the other curves. Silk, on the other hand, shows different behavior, with a drop for Case 1 of ~-10 dB at 10 kHz. Finally, Collagen-HA presents a significant deviation of the performance with drops of up to -10 dB.

These estimated positive/negative shifts are acceptable from a clinical perspective since they represent an amplification of the sound signal that might be beneficial to the patient, while a relative damping with the native middle ear occurs at high frequencies far from the speech frequency range.

Another interesting point of discussion is the material used to fabricate the prostheses. Although material choice does not significantly affect the mechanical performance of the prostheses, manufacturability may be a relevant issue, especially for non-conventional shapes.

Titanium-based prostheses with features in the order of microns can be made via electro-discharge machining (EDM) or 3D printing (Frazier, 2014). Recently, researchers have developed titanium prostheses that can be adjusted *ex situ* or *in situ* to allow a better coupling with the host tissues. However, only the *in situ* adjustable prostheses showed comparable acoustic performances to the fixed length prostheses, but they still require the interposition of a slice of cartilage to prevent rejection (Gottlieb et al., 2016). By choosing cortical bone as a constitutive material, the preferred approach is a fabrication with auto-/heterologous tissues from the bank, which requires the use of ultraprecision milling. However, the tiny dimensions of the devices and their features might be hard to achieve, also due to the brittleness of the bone itself (Milazzo et al., 2018, 2016). For its structural properties (D'Alessandro et al., 2012), compact bone is better used under a subtractive manufacturing approach; therefore, composite material alternatives, including collagen-based inks, have to be considered for additive manufacturing processing of ossicular bone devices (Milazzo et al., 2019).

Silk, on the other hand, is a natural protein fiber that can be easily used as an ink for Direct Ink Writing or, once solidified, shaped via ultraprecision milling. The good biocompatibility and tunable degradability of devices made of silk have made it a concrete alternative to synthetic materials for clinical applications (Li et al., 2016; Ling et al., 2016; Milazzo et al., 2019). Finally, Collagen/HA is a composite that resembles bone tissues as it contains the two most important constituents of bone. HA is the ceramic nanocomponent that strengthens the collagenous matrix. However, although the mineralization of cortical bone reaches about 70%, researchers have not overcome the upper limit of 40% due to embrittlement of the material (Jung and Buehler, 2017). Collagen/HA has been successfully shaped through molding and Direct Ink Writing, also embedding living cells into the ink with proper precautions, to create replacements for tissues with features in the order of millimeters (Milazzo et al., 2019).

Nevertheless, to the best of the authors' knowledge, applications in otology have not been pursued so far.

From a practical point of view, reaching such small features (especially for Case 4/L=7 mm) may represent a challenge. However, based on the material, we will investigate different approaches to fabricate the topologies as-they-are, with their optimized shape. For instance, as recently discussed (Milazzo et al., 2019), we could move from direct to indirect 3D printing (viz., molding with dedicated 3D printed molds able to deal with possible undercuts).

In case of specific manufacturability problems due to either material properties or fabrication procedures, a further simplification of the geometries may be necessary (e.g., as Case 4/L=7 mm, a filled-hole topology similar to Case 3/L=7 mm could be considered as optimal) looking for a trade-off between acoustic performances and practicability of the fabrication approach.

From the clinical perspective, the weight of the prostheses should not overcome the ossicular chain weight ($\approx$58 mg) for granting stability acoustic and outcomes. Generally, the lower is considered the better. This requirement is fully satisfied since the optimized prostheses weigh at the most 10 mg and 5 mg for titanium and cortical bone, respectively, and 3 mg for silk and Collagen/HA. Furthermore, the use of bone, silk, and Collagen/HA is preferred over titanium as they show a better biocompatibility. Indeed, titanium, which is largely employed in clinics, requires the interposition of a slice of cartilage between the prosthesis and the eardrum to reduce the risk of rejection, which occurs in the 20% of the cases in the long term. Moreover, placing a buffer between the tympanic membrane and the prosthesis footplate modifies the acoustic chain, which inevitably affects the acoustic transmission.

## 4. CONCLUSIONS

We present a study for the topology optimization and model validation of total ossicular replacement prostheses. Based on input from clinicians and data from the literature, we show how it is possible to maximize the stiffness of a middle ear prosthesis and to minimize the volume (and thus the weight) of the prosthesis, while ensuring the material continuity. Following earlier studies, we investigate how predefined holes on the umbo plate and their dimensions affect the final optimized geometry. Finally, through a finite-element model, including the eardrum and the stapedial footplate with annular ligament, we show that the prostheses behave similarly to the native ossicular chain under dynamic sound-pressure loading of the eardrum.

This work proposes a new approach to design middle ear prostheses with topologies not constrained to the classical geometries, which can be fabricated employing new techniques such as additive manufacturing or ultraprecision machining, which do not face the limitations experienced by the traditional approaches. Moreover, due to the linear regime in which such prostheses work, it is possible to employ different biocompatible materials (e.g., silk, Collagen/HA) in the optimization process without expecting significant modifications to the overall topology.

As a next step, it is important to fabricate the obtained designs by using 3D printing, and experimentally test this new generation of middle ear prostheses in human temporal bones. Furthermore, best fabrication parameters associated with each material should be investigated to achieve the highest precision and repeatability in the production process.

Finally, such a scientific and technologic advancement could also be used in other applications of tissue engineering where precision and reliability are required to create tissue replacements.

## CONFLICT OF INTEREST

All the authors declare no conflict of interest.

## ACKNOWLEDGMENTS

The authors want to thank Prof. Luca Bruschini (University of Pisa, Italy) and Dr. Andrea De Vito (Azienda Ospedaliero-Universitaria Pisana, Italy) for their precious suggestions on the clinical perspective.

This work has received funding from the European Union Horizon 2020 research and innovation program under the Marie Skłodowska-Curie grant agreement COLLHEAR # 794614. M.J.B. acknowledges support from NIH U01HH4977, U01EB014976, and U01EB016422. P.G.G.M. acknowledges support from the Research Foundation–Flanders (FWO), grant no. 11T9318N.

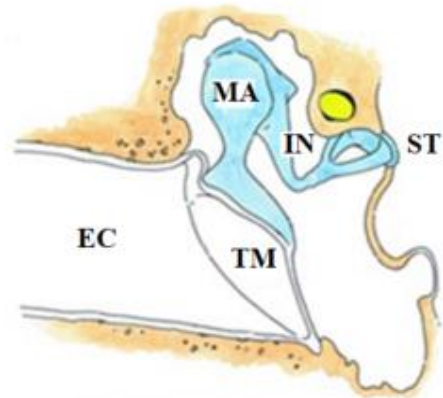
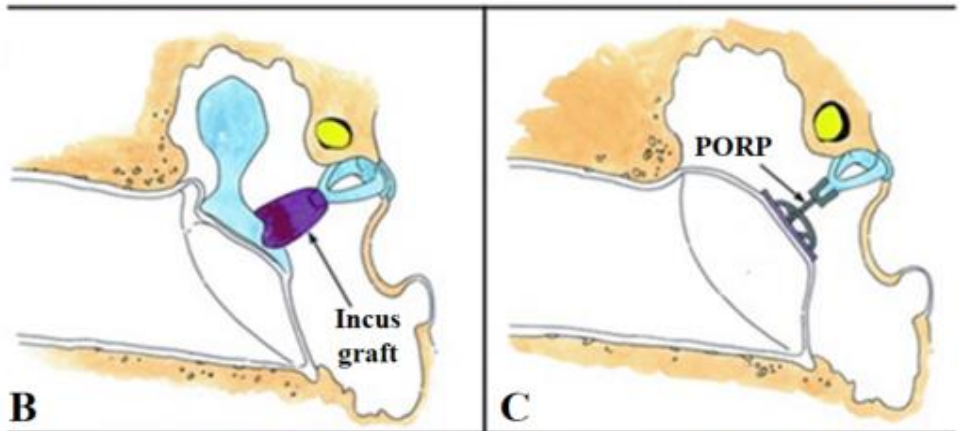
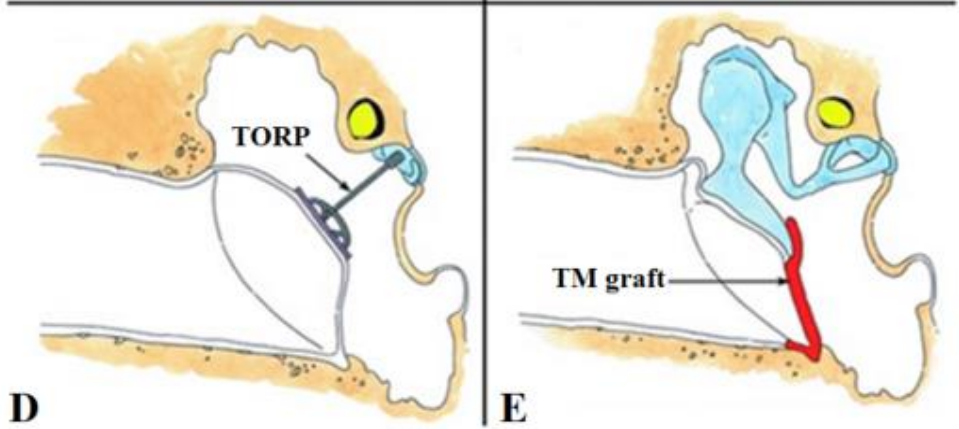

Fig. 1. Middle ear anatomy and main clinical solutions to repair conductive hearing loss. A. Healthy middle anatomy with legend of the main depicted parts. B. Incus graft to ensure material continuity of the ossicular chain. C. Partial Ossicular Replacement Prostheses (PORP), used in the presence of an intact stapes. D. Total Ossicular Replacement Prostheses (TORP), employed when the whole ossicular chain is removed. E. Tissue graft to replace the tympanic membrane, partially or totally. Adapted with permission. (Merchant and Rosowski, 2013) Copyright Springer Science+Business Media New York 2013.

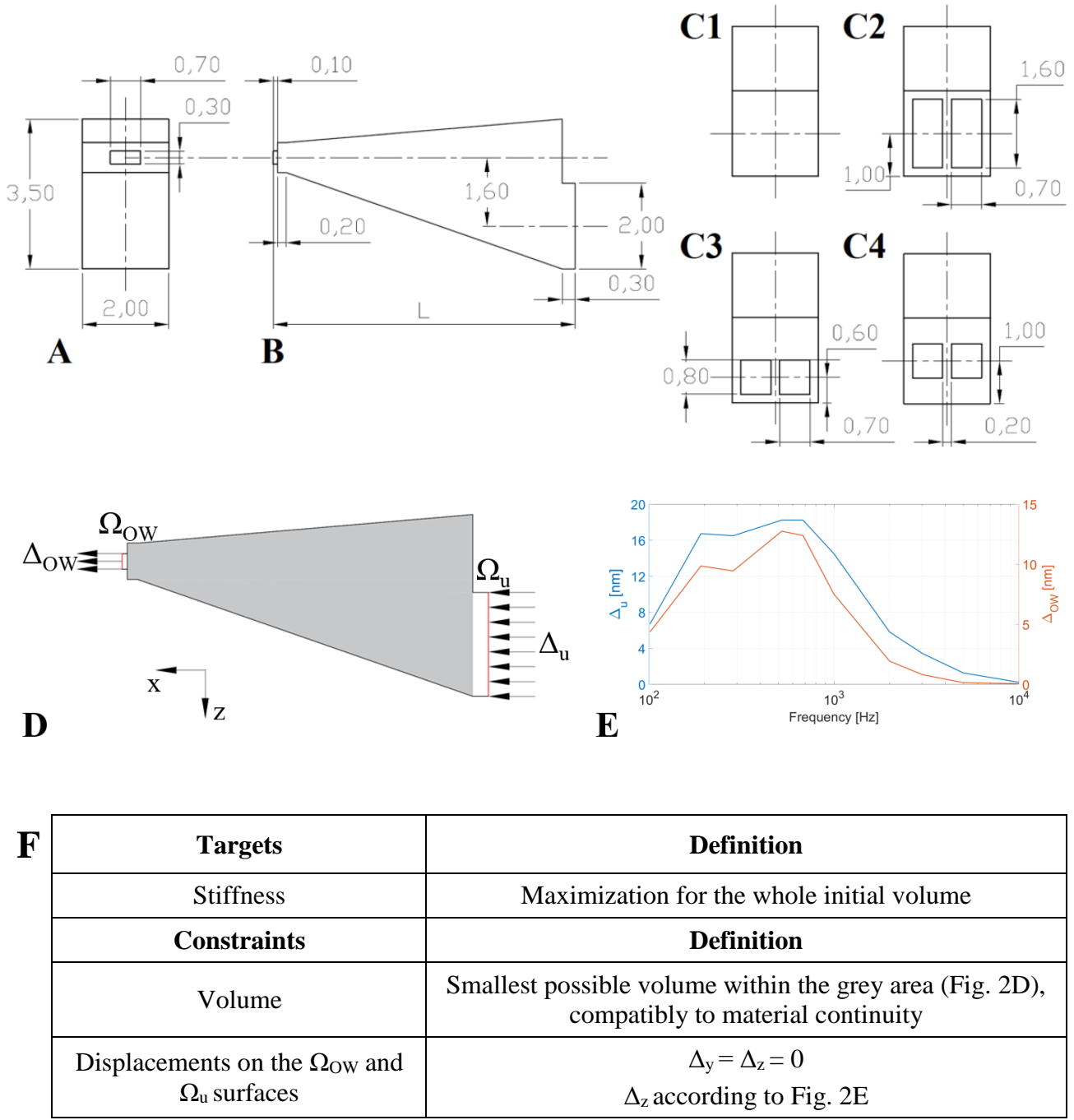

Fig. 2. Main dimensions, loads, and constraints of the body used as baseline for the topology optimization. A. The prosthesis viewed from the oval-window side, showing the interface with the oval window. B. Side view. Due to specific clinical needs, the length of the prosthesis (*L*) was set to 5 mm, 6 mm and 7 mm. C1-C4. Depiction of the four different umbo plate topologies: bulk (Case 1 – C1), big holes with borders of 0.2-mm thickness (Case 2 – C2), holes of half the height, asymmetrically positioned with the umbo plate (Case 3 – C3), and symmetrically positioned (Case 4 – C4). D. Schematic representation of the displacement loads applied to the surfaces in contact with the umbo ($\Omega_u$) and oval window ($\Omega_{OW}$). This two surfaces marked in red are fixed along the *y-/z-* directions. The bulk body subjected to topology optimization is colored in grey. E. Axial displacement as a function of frequency at the umbo and oval window in humans, measured by (Hato et al., 2003). F. List of the targets and constraints used for the optimization study.

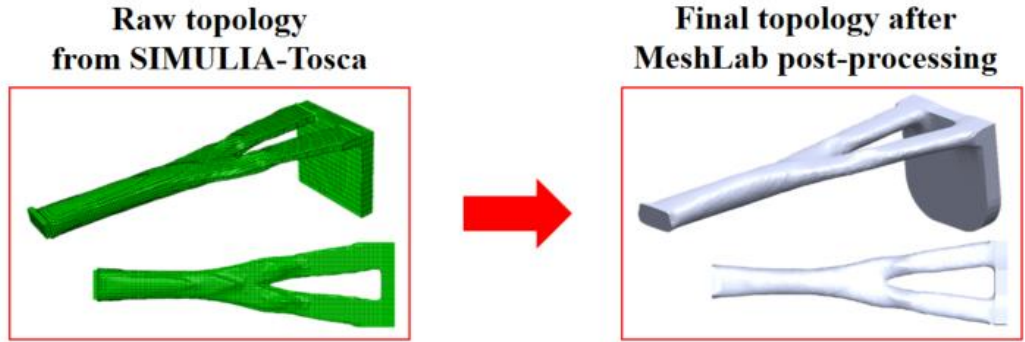

Fig. 3. Results from the topology optimization process for different prosthesis lengths (*L*) made of titanium. A. Post-processing of raw outcomes from SIMULIA-Tosca (green textures, left) leads to the final shape of the prosthesis after filtering and smoothing operations with MeshLab (grey textures, right). B. Prosthesis topologies using different lengths for the analyzed four cases. The 5-mm devices present the simplest piston-like geometry that does not significantly change with the presence and arrangement of the holes. The 6-mm and 7-mm prostheses differ from the 5-mm shapes, as they contain two/three branches that connect the umbo plate with the ellipsoidal shaft. Between the 6-mm and 7-mm prostheses, we observe similar results for Cases 1 to 3 with a relevant deviation for Case 4, in which we notice two flexible branches for the 7-mm device that are absent in 6-mm prosthesis.

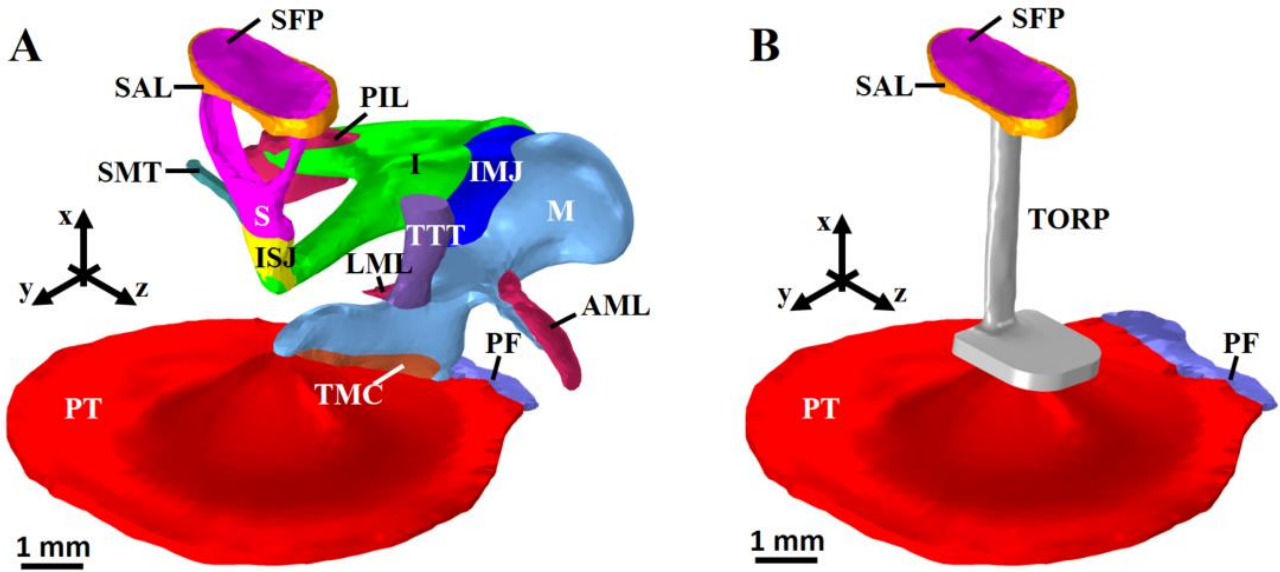

Fig. 4. Finite-element model employed to assess the performances of the middle ear prostheses. Panel A depicts the healthy middle ear according to (De Greef et al., 2017) while Panel B shows the adaptation of the model to our study, where a 5-mm TORP replaces the native middle ear components between the eardrum and oval window. Legend: PS: pars tensa, PF: pars flaccida, M: malleus, I: incus, S: stapes, IMJ: incudomalleolar joint, ISJ: incudostapedial joint, TMC: tympano-mallear connection; AML: anterior mallear ligament, LML: lateral mallear ligament, PIL: posterior incudal ligament, SAL: stapedial annular ligament, SFP: stapes footplate, TTT: tensor tympani tendon, SMT, stapedial muscle tendon.

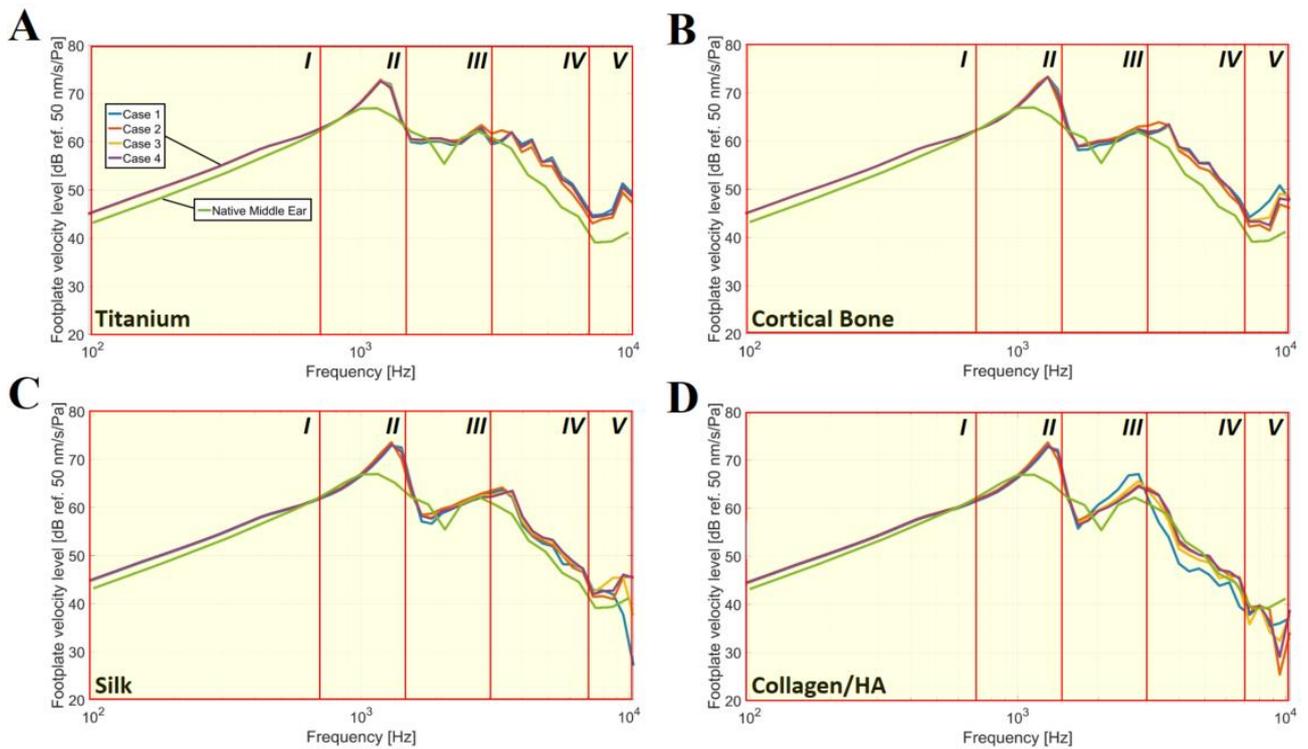

Fig. 5. Semi-logarithmic plots showing the footplate velocity level as a function of frequency for the native ossicular chain (green curve) compared with the performance evaluated for the four 5-mm prostheses (colored curves) made of titanium (Panel A), cortical bone (Panel B), silk (Panel C), and Collagen/HA (Panel D). We observe a general behavior for all prosthesis responses, which retrace the response of the native middle ear well. Only silk and Collagen/HA present a drop up to 10 dB in region V, above 7 kHz that, however, is acceptable since it this frequency range is far from the speech range (1-2 kHz).